\documentclass[twocolumn,english,10pt,journal,letterpaper]{IEEEtran}
\usepackage[T1]{fontenc}
\usepackage{color}
\usepackage{babel}
\usepackage{array}
\usepackage{float}
\usepackage{multirow}
\usepackage{amsmath}
\usepackage{graphicx}
\usepackage[none]{hyphenat}
\usepackage{amsthm, amssymb, amsmath}
\usepackage{braket}
\usepackage{color}
\usepackage{cleveref}
\usepackage{comment}
\usepackage{subcaption}
\usepackage{tabularx}

\usepackage{array}
\usepackage{ragged2e}
\newcolumntype{L}[1]{>{\RaggedRight\hspace{0pt}}p{#1}}

\usepackage[normalem]{ulem}

\makeatletter

\renewcommand\footnoterule{ 
  \kern-3\p@
  \hrule\@width 5cm
  \kern2.6\p@}
  
\def\subsection{\@startsection{subsection}{2}{2.5\parindent}{1.5ex plus 1.5ex minus 0.5ex}%
{0.7ex plus .5ex minus 0ex}{\normalfont\normalsize\itshape}}%

\makeatother

\begin{document}

\renewcommand{\figurename}{Fig.}
\renewcommand{\tablename}{TABLE}

\title{Leveraging Bitcoin Mining Machines in Demand-Response Mechanisms to Mitigate Ramping-Induced Transients}

\author{Elinor~Ginzburg-Ganz, Ittay~Eyal, Ram~Machlev, Dmitry~Baimel, Leena~Santosh, Juri~Belikov, Yoash~Levron

\thanks{The work of J. Belikov was partly supported by the Estonian Research Council grant PRG1463.}

\thanks{E.~Ginzburg-Ganz, R. Machlev, and Y. Levron are with the Andrew and Erna Viterbi Faculty of Electrical \& Computer Engineering, Technion---Israel Institute of Technology, Haifa, 3200003, Israel (elinor.g12@gmail.com, rmhalb@gmail.com, yoashl@ee.technion.ac.il). I. Segev and S. Keren are with Computer Science Faculty of Electrical and Computer Engineering, Technion---Israel Institute of Technology, Haifa, 3200003, Israel (itaysegev@campus.technion.ac.il, sarahk@technion.ac.il). D. Baimel is with the Shamoon College of Engineering, Beer-Sheva 84100, Israel (dmitrba@sce.ac.il). L. Santosh is with Pandit Deendayal Energy University (leena80.santosh@gmail.com). J. Belikov is with the Department of Software Science, Tallinn University of Technology, Akadeemia tee 15a, 12618 Tallinn (juri.belikov@taltech.ee).}

\thanks{Paper submitted to the International Conference on Power Systems Transients (IPST2025) in Guadalajara, Mexico, June 8-12, 2025.}

}

\maketitle
\begin{abstract}
We propose an extended demand response program, based on ancillary service for supplying flexible electricity demand. In our proposed scheme, we suggest a broader management model to control the scheduling and power consumption of Bitcoin mining machines. The main aspect that we focus on is suppressing the power ramping and related transient effects. We extend previous works on the subject, that study the impact of incorporating cryptocurrency mining machines into existing power grid, and explore the potential profit of exploiting this flexible load in the Israeli electricity market. We analyze a trend based on historical data, of increasing electricity prices and ramping costs due to the increasing penetration of renewable energy sources. We suggest an extension to the unit commitment problem from which we obtain the scheduling scheme of the Bitcoin mining machines. We use simulation and the real-world data acquired from the ``Noga’’ grid operator to verify the proposed ancillary service and test its practical limits for reducing the ramping costs, under changing ratio of energy production from renewable sources. Out results suggests that the machine price and ratio of production from renewable sources plays a significant role in determining the profitability of the proposed demand-response program
\end{abstract}

\begin{IEEEkeywords}
Bitcoin, Demand response, Energy storage, Power demand, Power system transients,  Renewable energy sources 
\end{IEEEkeywords}

\section{Introduction}

\IEEEPARstart{R}{amping}, or the ability of power systems to rapidly adjust generation output to match sudden fluctuations in electricity demand or supply, presents a significant challenge in modern power systems, particularly with the increasing integration of renewable energy sources. Unlike conventional power plants that can provide consistent power output, renewables like wind and solar are inherently variable, leading to rapid changes in power generation that can be difficult to predict and manage. These fluctuations may induce transient phenomena, which cover can significantly impact the stability and performance of the grid. These transients, occurring due to rapid shifts in power generation or load, can lead to voltage sags, swells, or even blackouts if not properly managed. As the penetration of variable renewable energy sources continues to grow, alongside the growing demand, the frequency and intensity of these transient events increase, posing substantial risks to the reliability and resilience of the power grid. High power peaks, or sudden surges in electricity demand, place immense pressure on power generation and distribution infrastructure, often during specific times of the day, such as early evenings when residential consumption spikes.
To ensure reliability and avoid blackouts, maintaining a balance between supply and demand is essential. However, simply building additional power plants or upgrading the grid to handle these peaks is both costly and environmentally unsustainable. Demand-response programs offer a more sustainable solution by adjusting demand rather than increasing supply, encouraging consumers to reduce or shift their energy usage during peak periods. This approach not only helps stabilize the grid but also promotes energy efficiency and reduces the need for costly infrastructure upgrades.

In addition to traditional demand-response strategies, utilizing massive power consumers like Bitcoin mining machines presents a novel approach for balancing the grid demand. These machines are significant power consumers, and may be operated when there is a need to increase power usage during periods of low energy consumption, to abstain from powering off generators and reducing ramping costs, or when there is an abundance of renewable energy generation, to avoid energy curtailment. By doing so, they help substantially reduce ramping costs and absorb excess energy that might otherwise be wasted, providing a flexible demand source that can be dialed up or down based on grid conditions very quickly. Moreover, the revenue from mining is another incentive to use these machine in for enhancing grid stability, instead of other proposed solution such as storage or curtailment. The profits from these machines might cover their operational costs, thus resulting in an economically feasible solution for power plant operators.
This not only maximizes the utilization of renewable energy but also helps maintain grid stability by smoothing out fluctuations in demand and supply, making the overall power system more resilient and efficient.

In recent literature, many researchers study this problem from different perspectives. For instance, work \cite{Yaghmaee2024} evaluates the profitability of Bitcoin miners in terms of energy prices, by a control algorithm that schedules miners based on the energy price, Bitcoin price, and total network hash rate. The authors propose a demand response program to control the power load in the network and balance supply and demand. The demand response program is modeled as a mixed integer linear programming optimization problem. An extensive review is then presented in \cite{Mololoth2023}. Here, they examine the potential of combining blockchain technology and machine learning techniques in the development of smart grids and the benefits achieved by using both techniques. Work \cite{Kumari2022} focuses on smart grid systems necessitating secure demand response management schemes for real-time decision-making to increase the effectiveness and stability of smart grid systems along with data security. The authors propose a secure DRM scheme for home energy management using Q-learning algorithm and Ethereum blockchain to make optimal price decisions when the objective is to reduce energy consumption and decrease costs. In addition, \cite{Jamil2021} concerns with peer-to-peer energy trading. In this paper, the writers propose a blockchain-based predictive energy trading platform to provide real-time support, day-ahead controlling, and generation scheduling of distributed energy resources. This study aims to achieve optimal power flow and energy crowd-sourcing, supporting energy trading among the consumer and prosumer. Moreover, paper \cite{Karandikar2021} addresses the problem of peak shaving. In the article, they design and implement a blockchain-based prosumer incentivization system, in which the smart contract logic is based on the in-depth analysis of the ``Ausgrid'' dataset. Furthermore, \cite{Samadi2021} presents a demand response solution utilizing energy blockchain to reduce demand, in order to save the extra Distributed Energy Resources (DER), and efficiently incorporate customers' block mining ability. The authors deploy a Stackelberg game between a control agent and local customers to negotiate demand reduction by integrating the block mining method as DERs saving. The solution is an innovative consensus algorithm, ``proof of energy saving'', that is used to incentivize the customers to reduce their demand, discharge their electric vehicle, and maximize their chance for block mining to earn monetary rewards and DERs savings. Additionaly, \cite{Cioara2018} proposes a demand response framework for near-real-time autonomous demand response management combined with a market-driven pricing scheme. The approach aims to aggregate groups of prosumers and develop new financial and business models for maximizing the prosumer's benefits in terms of renewable energy usage maximization, and cost minimization. Finally, \cite{Shifan2021} is concerned with a demand response program aimed at optimizing energy usage in buildings. They approach the problem using a non-cooperative game theory framework and implement a customized blockchain-based energy optimization schedule they developed.

\textbf{Contribution}: As shown in the literature review above, recent works have already proposed strong methods for operating mining machines (for Bitcoins, or other cryptocurrencies) for regulating the load and enhancing the grid stability. These methods typically follow the usual patterns of demand-response mechanisms, where generally the lower the total load, the more power is supplied to mining machines, and vice-versa \cite{Karandikar2021, Menati2023Modeling}. Previous works have investigated many aspects of this problem, and investigated for instance the profitability of these machines in the context of demand-response, and the use of smart contracts when renewable energy sources are included in the generation mix. In this paper we extend these previous works by including the aspect of ramping constraints, which are crucial in power systems with a high penetration level of renewable energy sources, as demonstrated by the famous ``duck curve''. In comparison to previous works, we assume that machine owners are compensated not only for the generated power and lowering the power peak, but also for mitigating fast ramps, which are problematic for the system operator, thus
we explore an optimization problem that takes into account the electricity costs for the grid operator, the ramping costs, and the revenue from the machines. We show that this problem can be efficiently solved based on Pontryagin's minimum principle, and thoroughly explore the profitability of the machines in this scenario. Our simulations are based on data taken from the ``Noga'' grid operator, which is examined under changing ratios of energy production from renewable sources. In our analysis, we compare the effects of several key parameters, such as the electricity price, and the machines' price, the machines' hashrate and monetary revenue. These are examined for several different machine types that are available in the market today. Our main conclusion is that the profitability of the discussed mechanism is highly influenced by the cost of the mining machines, and the percentage of renewable sources within the energy mix, where some scenarios are more profitable then others.

\section{Main Result}
We consider a power system operator that attempts to balance its load demand using cryptocurrency mining machines (for instance, Bitcoin mining machines), which will be referred to as ``miners''. Our simplified model consists of a grid-connected mining machine and an aggregated load.
The load is described by its active power demand, which is represented by the function $p_{L}(t): \mathbb{R}_{\geq 0} \to \mathbb{R}_{\geq 0}$ over a finite time interval $[0,T]$ for some given and known time $T$. 
The power supplied by the generator is denoted by $p_{g}(t): \mathbb{R}_{\geq 0} \to \mathbb{R}_{\geq 0}$ and is associated with a cost function of the fuel consumption $c_{g}(t)$. More generally, this cost function may represent a general cost function with various objectives, such as carbon emission influenced by power generation. The miner is characterized by its power demand $0 \leq p_{m}(t) \leq \bar{P}$, and the profit obtained by its operation is $c_{m}(t)$. We assume the miner's power consumption is much smaller than the total system's mining power (all Bitcoin miners), thus, it is justified to use a cost function $c_{m}(t)$ which is linear in the miner's power consumption. The instantaneous cost of electricity is
\begin{equation}
    F(t) = c_g(t)p_g(t) + c_d(t) \left( \frac{\text{d}p_g}{\text{d}t} \right) - c_m(t)p_{m}(t).
\end{equation}
This cost reflects (a) the fuel cost associated with the generated power, (b) the cost associated with rapid changes in generated power, which is represented by the term $c_d(t)\frac{\text{d}p_g}{\text{d}t}$, and (c) the revenue of the grid operator from using the mining machines. The objective of the grid operator is to minimize the total cost $\int_{0}^{T}{F(t)\text{d}t}$ by choosing the optimal function $p_{m}(t)$, where the time horizon $T$ is known. This leads to the following optimization problem:
\begin{equation}
\begin{aligned}
    \min_{\{ p_{m}(\cdot) \}} & \int_{0}^{T}{\left( c_g(p_g(t)) + c_d \left(\frac{\text{d}p_g}{\text{d}t} \right) - c_m(t)p_{m}(t)  \right) \text{d}t}, \\
    \text{s.t. } & p_g(t) = p_{L}(t) + p_{m}(t), \\
    & 0 \leq p_m(t) \leq \bar{P}, 
\end{aligned}
\label{eq:optimization-problem}
\end{equation}
where all the functions and constants are known and given, and the decision variable is the function~$p_{m}(\cdot)$. It is assumed that $c_{g}(\cdot), c_{d}(\cdot) \in \mathcal{C}^{2}$ are strictly convex functions, and the derivatives of $c_{g}, c_{d}$, denoted as $c'_{g}, c'_{d}$, defines a mapping from~$\mathbb{R}$ to~$\mathbb{R}$, and we assume that $p_{L}(t), c_{m}(t)$ are smooth periodical functions with a period $T$. We use the function~$\xi$ to eliminate the inequality constraint,
\begin{equation}
    \xi(p_{m}) = \begin{cases}
        0, \text{ for } 0 \leq p_{m} \leq \bar{P}\\
        \alpha p_{m}^2, \text{ for } p_{m} < 0 \\
        \alpha (p_{m} - \bar{P})^2, \text{ for } p_{m} > \bar{P} \\
    \end{cases}
\end{equation}
where $\alpha$ is a constant.

The new formulation that arises is,
\begin{equation}
\begin{aligned}
    \min_{\{ p_{m}(\cdot) \}} & \int_{0}^{T}{c_g(p_g(t)) + c_d \left(\frac{\text{d}p_g}{\text{d}t} \right) - c_m(t)p_{m}(t) + \xi(p_{m}) \text{d}t}, \\
    \text{s.t. } & p_g(t) = p_{L}(t) + p_{m}(t), \\
    & \xi(p_{m}) = \begin{cases}
        0, \text{ for } 0 \leq p_{m} \leq \bar{P}\\
        \alpha p_{m}^2, \text{ for } p_{m} < 0 \\
        \alpha (p_{m} - \bar{P})^2, \text{ for } p_{m} > \bar{P} \\
    \end{cases}
\end{aligned}
\end{equation}
when $\alpha$ approaches infinity, this last formulation is equivalent to \eqref{eq:optimization-problem}.

There are many approaches for solving this category of optimization problems, 
one such approach uses Pontryagin's Minimum Principle \cite{HannemannTamas-2012-How}. Applying this principle, we define the following:
\begin{equation}
    \begin{aligned}
        x(t) & = p_{g}(t), \\
        u(t) & = \frac{\text{d}}{\text{d}t} p_{g}(t), \\
        \Omega & = [-\infty, \infty], \\
        t_1 & = T, \\
        f(x,u) & = u, \\
        L(x,u,t) & =  c_g(x) + c_d \left(u \right) - c_m(t) \left(x - p_L(t) \right) \\
        & + \xi \left(x - p_L(t) \right).
    \end{aligned}
\end{equation}

Let us define the Hamiltonian
$$H(x, \lambda, u, t) = L(x,u,t) + \lambda f(x,u).$$
Using Pontryagin's minimum principle, the necessary conditions for an optimal solution $x^{*}(t), u^{*}(t), \lambda^{*}(t)$ to exists, are the following ones:
\begin{enumerate}
    \item $\frac{\text{d}}{\text{d}t} x^{*}(t) = u^{*}(t),$
    \item $\frac{\text{d}}{\text{d}t} {\lambda}^{*} = -c'_g(x^{*}) + c_m(t) - \xi'(x^{*} - p_L(t))$,
    \item $H(x^*, \lambda^*, u^*, t) \leq H(x^*, \lambda^*, u, t)$ for all $u$.
    \item $x^*(0) = x^*(T)$, and $\lambda^*(0)=\lambda^*(T)$. This is because $p_L(t), c_m(t)$ are periodical, and so the optimal solution must be periodical as well.
\end{enumerate}
Where $c'_g(\cdot)$ is the derivative of $c_g(\cdot)$, and $\xi'(\cdot)$ is the derivative of $\xi(\cdot)$. The third condition is
\begin{equation}
    c_{d}(u^*)+ \lambda^{*} u^{*} \leq c_{d}(u)+ \lambda^{*} u \quad \text{ for all } u.
\end{equation}
Since both sides of the equation are convex, the optimal $u^*$ can be found by zeroing the derivative:
\begin{equation}
    c'_{d}(u^*)+ \lambda^{*} = 0.
\end{equation}
According to our assumption, $c'_{d}(\cdot)$ is strictly monotonically increasing, and defines a mapping from $\mathbb{R}$ to $\mathbb{R}$. Therefore, we can write
\begin{equation}
    u^* = (c'_{d})^{-1}(-\lambda^{*}).
\end{equation}
This leads to the following explicit conditions:
\begin{enumerate}
    \item $\frac{\text{d}}{\text{d}t} x^{*}(t) = u^{*}(t)$,
    \item $\frac{\text{d}}{\text{d}t} {\lambda}^{*} = -c'_g(x^{*}) + c_m(t) - \xi'(x^{*} - p_L(t))$,
    \item $u^* = (c'_{d})^{-1}(-\lambda^{*})$,
    \item $x^*(0) = x^*(T)$, and $\lambda^*(0)=\lambda^*(T)$.
\end{enumerate}

To show an explicit solution for a simple case, assume that $\xi'(x^{*} - p_L(t)) = 0$, $c_{d}(z) = a z^2$ and $c_m(t) = c_m =$ const. In that case, the optimal solution is
\begin{equation}
    x^{*}(t) = (c'_g)^{-1}(c_m), \quad u^{*} = 0, \quad \lambda^{*}=0.
\end{equation}

\section{Numerical Results}
We examine the opportunity of incorporating Bitcoin mining machines as part of the operation of ``Noga'' grid operator to reduce ramping charges, and investigate how the profitability of the machines is affected by renewable energy penetration to the market, as electricity prices change. Further, we examine realistic properties of the Bitcoin mining machines, in terms of machine price, their power consumption, and hashrate, to ensure maximal profitability for the power plant operator. The machine price directly affects the initial investment and payback period, while power consumption influences the ongoing operational costs, especially in energy-intensive mining processes. Hashrate, a measure of computational power, determines the machine's efficiency in solving cryptographic puzzles, and earning rewards. Together, these factors are essential for evaluating the assumed profitability of mining operations in relation to the power plant's resources.

First, we assess how the following factors influence the profitability of using Bitcoin mining machines for regulating demand-response in power plants, and decreasing ramping costs: (1) The influence of the ratio of renewable energy production out of the total production on electricity costs and thus on the profits gained from mining machines. (2) The influence of the ratio of renewable energy production out of the total production on ramping costs and thus on the profits gained from mining machines. Next, we evaluate the potential profitability of using Bitcoin mining machines by the power company ``Noga'' for mitigating ramping effects and consuming excess power that is generated in the Israeli grid. In recent years, the integration of renewable energy resources, especially those based on solar energy, has evolved considerably in this country, introducing a major problem of uncontrolled excess energy production, which jeopardizes the operation of the power grid. Since Israel is considered an energetic island, in case the production is higher than the electricity the grid can absorb, particularly during periods of low demand and high generation, the excess electricity may be treated in the following ways: (1)~Routing to large storage systems, such as ``Ashalim'', which are located in sparsely populated areas in the south of Israel. Unfortunately, the transmission infrastructure has relatively low capacity~\cite{Gal2020}, leading to substantial losses; (2)~Reducing production, which has heavy monetary implications, partially due to ramping costs~\cite{Wang1995}; (3)~It may be curtailed. Each option has its disadvantages

Now we proceed to examine several machine parameters, focusing on machine prices, power consumption, and machine hash rate, upon which routing the excess energy during off-peak hours to the Bitcoin mining machines, will not only reduce ramping costs, and help sustain the longevity of infrastructure, but also produce profit to the machine operators. For each parameter set, we provide the optimal operation scheduling for different types of loads and renewable energy production in percentage, while trying to find the optimal parameter set, which will produce revenue for the system operator for the maximal time horizon, when taking into account the growing percentage of renewable energy production, changing electricity costs, and profit decay of the machines.

\begin{figure}[htbp]
    \centering
    \includegraphics[width=\linewidth]{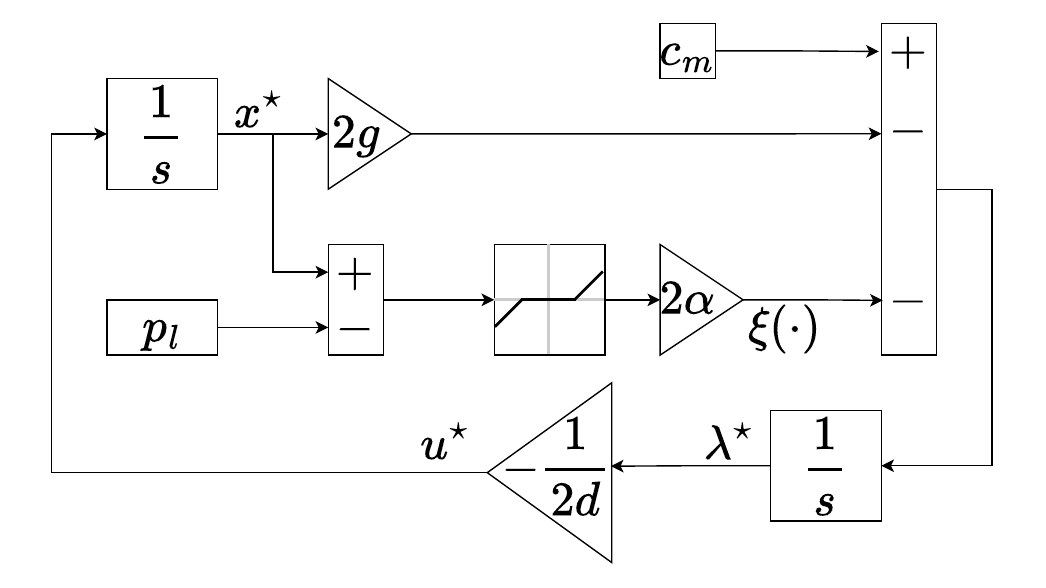}
    \caption{A control loop that implements the optimal scheduling of the miner, as derived from Pontryagin's Minimum Principle.} 
    \label{fig:pontryagin-solution}
\end{figure}

\begin{figure}[htbp]
    \centering
    \includegraphics[width=0.5\linewidth]{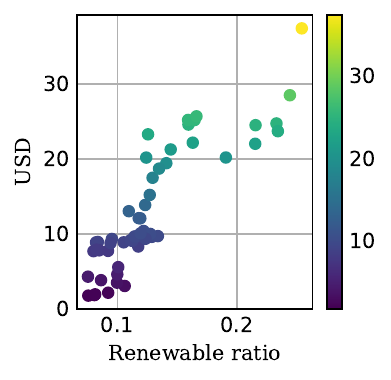}
    \caption{Renewable energy production in \% out of overall production effect on electricity prices in the state of California between the years 1970-2022.}
    \label{fig:renewable-production-energy-prices}
\end{figure}

We start by analyzing the effect of renewable energy production, specifically from solar sources, on the electricity prices. This data is not available for this grid operator, but it is available for the state of California, which has similar sun irradiation conditions. Hence, to perform this evaluation, we used historical data from the state of California, that includes renewable and primary production data, in addition to electricity prices, between the years 1970-2022 provided in \cite{eia}. 
In Fig.~\ref{fig:renewable-production-energy-prices}, one can easily observe that as more and more renewable energy sources penetrate into the market, and their share in the overall production increases, then electricity prices rise. This happens partially due to grid defection \cite{navon2024stability}, and also to account for the inertia and reactive power correction that must be supplied by power plant operators, to keep the stability of the grid \cite{navon2020integration}.

\begin{figure}[htbp]
    \centering
    \includegraphics[width=0.5\linewidth]{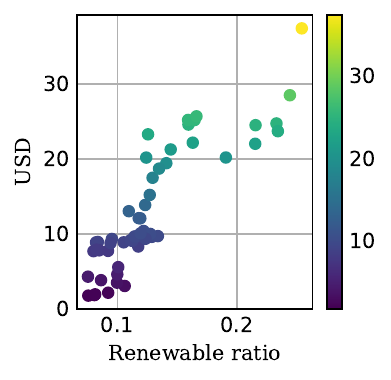}
    \caption{Bitcoin mining machines profitability in [\$], as a function of ratio of solar energy sources production.}
    \label{fig:mining-profitability-renewable}
\end{figure}

The profitability of Bitcoin mining machines is very sensitive to changes in electricity prices. Following this analysis, we may estimate how the increasing percentage of renewable energy penetration, such as the trend observed in California and Israel, affects the mining machines' profitability. For the simulation, we chose a few popular Bitcoin mining machines, their parameters are displayed in \Cref{tab:machine-parameters}, and they are based on data acquired from \cite{Bitmain}.
\begin{table}[htbp]
    \centering
    \begin{tabular}{|p{2cm}|p{1cm}|p{1cm}|p{1cm}|p{1.5cm}|}
        \hline
         Machine & Demand [W] & Hashrate [Th/s] & Income [\$/day] & Electricity costs [\$/day]\\
         \hline
         Antminer S21  & 5360 & 335 & 15 & 0.1 \\
         \hline
         Whatsminer M63 & 7283 & 334 & 14.49 & 0.1 \\
         \hline
         Antminer S19 & 3250 & 110 & 5.05 & 0.06 \\
        \hline
    \end{tabular}
    \caption{Bitcoin mining machines parameter sets taken from Bitcoin mining companies.}
    \label{tab:machine-parameters}
\end{table}
The dependency of Bitcoin mining machines profit upon the electricity costs is quite complex, and is extensively discussed in \cite{Delgado2019The}. However, in literature it is common to rely on a simplified model in which the monetary revenue of the Bitcoin mining machine is linear in the electricity price, for example, as seen in \cite{Tedeschi2024Mining}. In work \cite{Tedeschi2024Mining} the authors analyze a complex revenue model, accounting for marginal factors such as network hash rate, machine hash rate, and the time to mine a block, alongside with other factors including transaction fees and block reward, their hashing power and the probability of successfully mining a block. They show that the most influential factor on the profitability of a machine is the electricity prices, hence, giving the incentive to use a linear model for the relation between machine income and the electricity costs. Thus, the following relation describes the dynamics:
\begin{equation}
    \text{Profit} = a - b \cdot \text{Electricity-Price}.
\end{equation}
The parameter $a$ describes the maximal daily monetary profit that can be achieved, meaning, the daily monetary revenue from the Bitcoin mining machine when the electricity price is zero, and $b$ is a linear coefficient that represents how profit changes with electricity price. For our simulation we relied on realistic parameters based on the following work \cite{Tedeschi2024Mining} and standard machine properties acquired online from sites providing updated information on Bitcoin mining, such as \cite{coinDesk}. The values are $a=14$, and $b=0.1$. From Fig.~\ref{fig:mining-profitability-renewable}, it may be seen that there is a steep incline in the plot profile, meaning that the effect of renewable sources penetration is considerable when looking at mining machines' profits.

Now, as the second stage, we would like to present the trend of ramping costs as the percentage of renewable energy sources grows. The duck curve is a well-known problem that visually displays the escalation in ramping costs, as may be viewed in~\cite{duck-curve}.

\begin{figure}[htbp]
    \centering
    \includegraphics[width=0.5\linewidth]{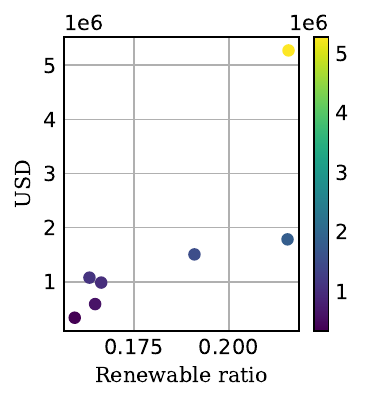}
    \caption{Renewable energy penetration effect on daily ramping costs in USD per Watt in California, based on historical data.}
    \label{fig:ramping-renewable-california}
\end{figure}

Figure~\ref{fig:ramping-renewable-california} displays the ramping costs, assuming a quadratic relation between the transient profile calculated in units of MW/h and the monetary value (based on \cite{quadratic-cost-function}), as a function of renewable energy production in percentage, out of overall production, is presented for the state of California. It is clear that there is a substantial escalation in ramping costs, as more renewable energy sources take a bigger chunk of the overall energy supply.

\begin{figure}[htbp]
    \centering
    \includegraphics[width=0.6\linewidth]{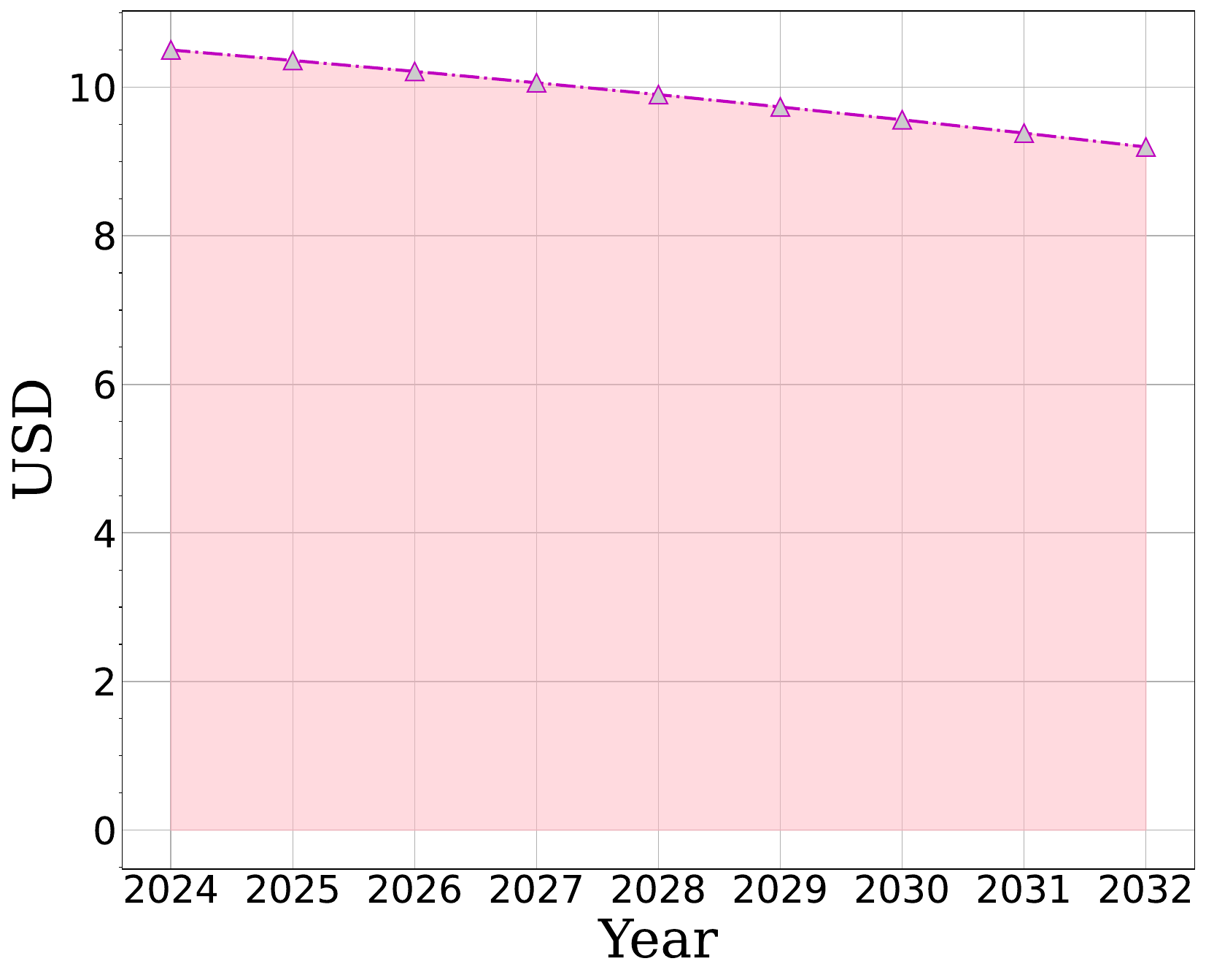}
    \caption{The operator's revenue from Bitcoin mining machines over time, due to renewable energy sources integration and plummeting electricity prices.}
    \label{fig:mining-profitability-years}
\end{figure}

Consequently, based on Figs.~\ref{fig:mining-profitability-renewable} and \ref{fig:ramping-renewable-california} it may be inferred that there is tension between rising electricity prices, which reduce the profitability of Bitcoin mining machines, and the increasing ramping costs, which incentivize the utilization of flexible and quick to respond consumers like Bitcoin mining machines. The idea is shown in Fig.~\ref{fig:mining-profitability-years}, which presents power plant operators revenue from mining, which takes into account a prediction of electricity prices and ramping costs based on historical data. It may be observed that when excluding machine prices, even for a modest revenue of $14$ USD it is profitable to utilize these mining machines in the next years, as the share of renewable sources increases.  

Building on the aforementioned results, we would like to determine optimal machine parameters that companies in Israel, managing power production, could use in their plants to mitigate ramping costs, and use profits from these Bitcoin mining machines to sustain their operation. Our simulation uses load and generation data acquired by ``Noga'' which are available to the public \cite{Noga}, normalized to represent a single power plant production. We chose realistic Bitcoin mining machine parameters based on three families of Bitcoin mining machines: ``Antminert-S19'', ``Antminert-S21 Pro'', and ``Whatsminerm63''. Since the parameters may change slightly between providers, we used estimated price and labeled the Bitcoin mining machines tested by the labels: ``1''-``3''. The parameters of the tested Bitcoin mining machines are presented in \Cref{tab:example-machine-parameters}.

\begin{table}[htbp]
    \caption{Bitcoin mining machines parameter sets}
    \label{tab:example-machine-parameters}
    
    \begin{tabular}{|p{0.5cm}|p{1.5cm}|p{1.4cm}|p{1.8cm}|p{1.5cm}|}
         \hline
         Type & Demand [W] & Hashrate [Th/s] & Income [\$/day] & Electricity costs [\$/W]\\
         \hline
         1 & 5360 & 335 & 15 & 0.1 \\
         \hline
         2 & 7283 & 334 & 14.49 & 0.1 \\
         \hline
         3 & 3250 & 110 & 5.05 & 0.06 \\
         \hline
    \end{tabular}
    
\end{table}

We relied on quadratic cost functions, which are well-known in literature \cite{quadratic-cost-function}. The cost function considered for power generation is given by $c_g(x) = g x^2$, where $g = k \cdot (\text{Electricity cost} / \text{Power consumption}^2)$ and $k$ is a constant, which has the value of $k=0.0014$ for machines number ``1'' and ``3'', and $k=0.0012$ for machine number ``2''. The cost of ramping is $c_d(x) = x^2$. The revenue per kWh is given by $c_m = \text{income} / \text{consumption}$, where the income, calculated in USD/day is based on average profits declared by miners and presented in sources such as \cite{coinDesk}. The consumption is defined by the machine power consumption over 24-hour time horizon: $\text{consumption} = \text{machine power consumption} \cdot 24$. The parameter of the function $\xi$ used for eliminating inequality constraints is $\alpha = 1$. The initial conditions are $x_0 = c_m / 2g$ and $\lambda_0 = 0$.

\begin{figure*}[htbp]
        \subfloat[October]{%
            \includegraphics[width=.4\linewidth]{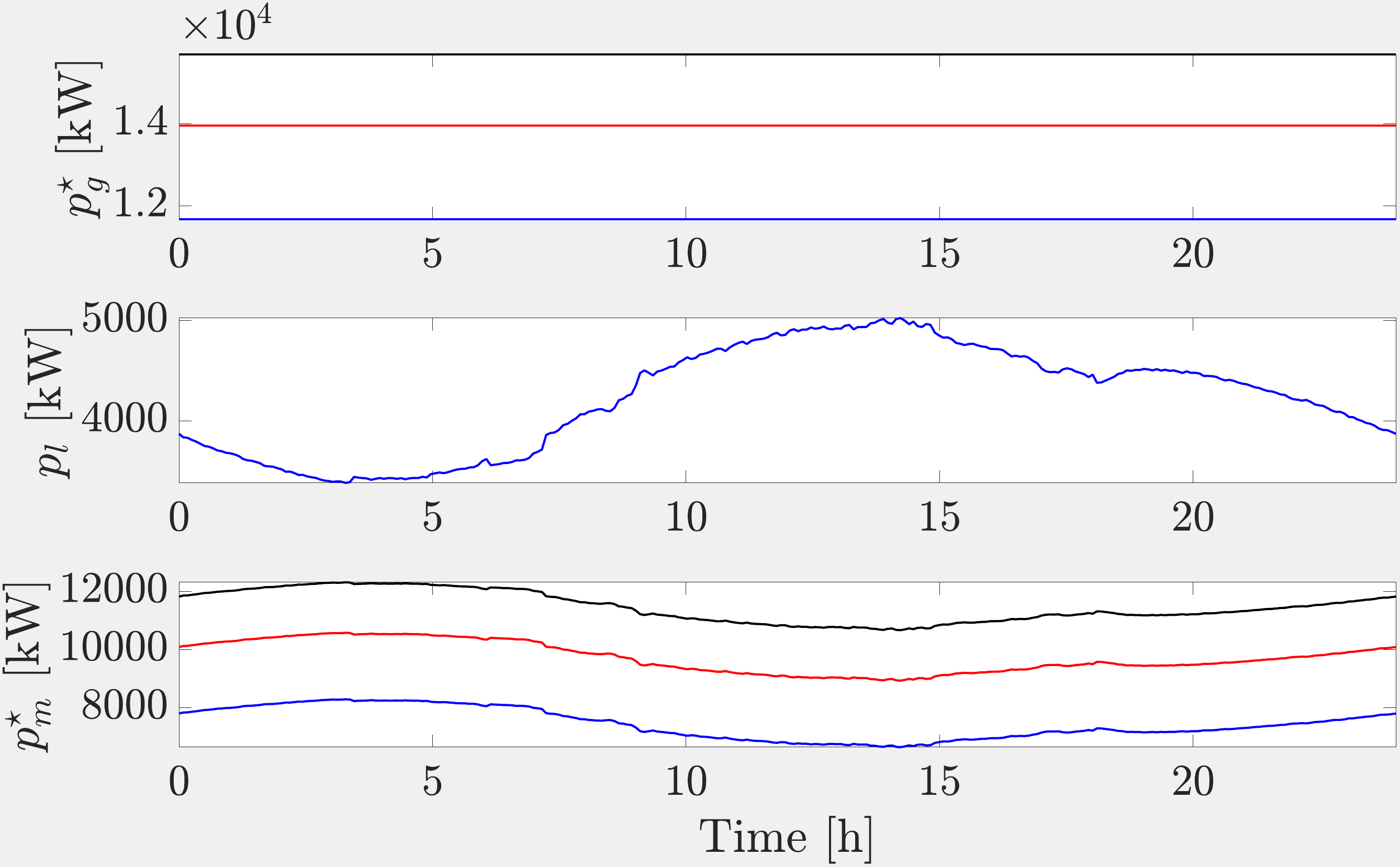}%
            \label{subfig:a}%
        }\hfill
        \subfloat[April]{%
            \includegraphics[width=.4\linewidth]{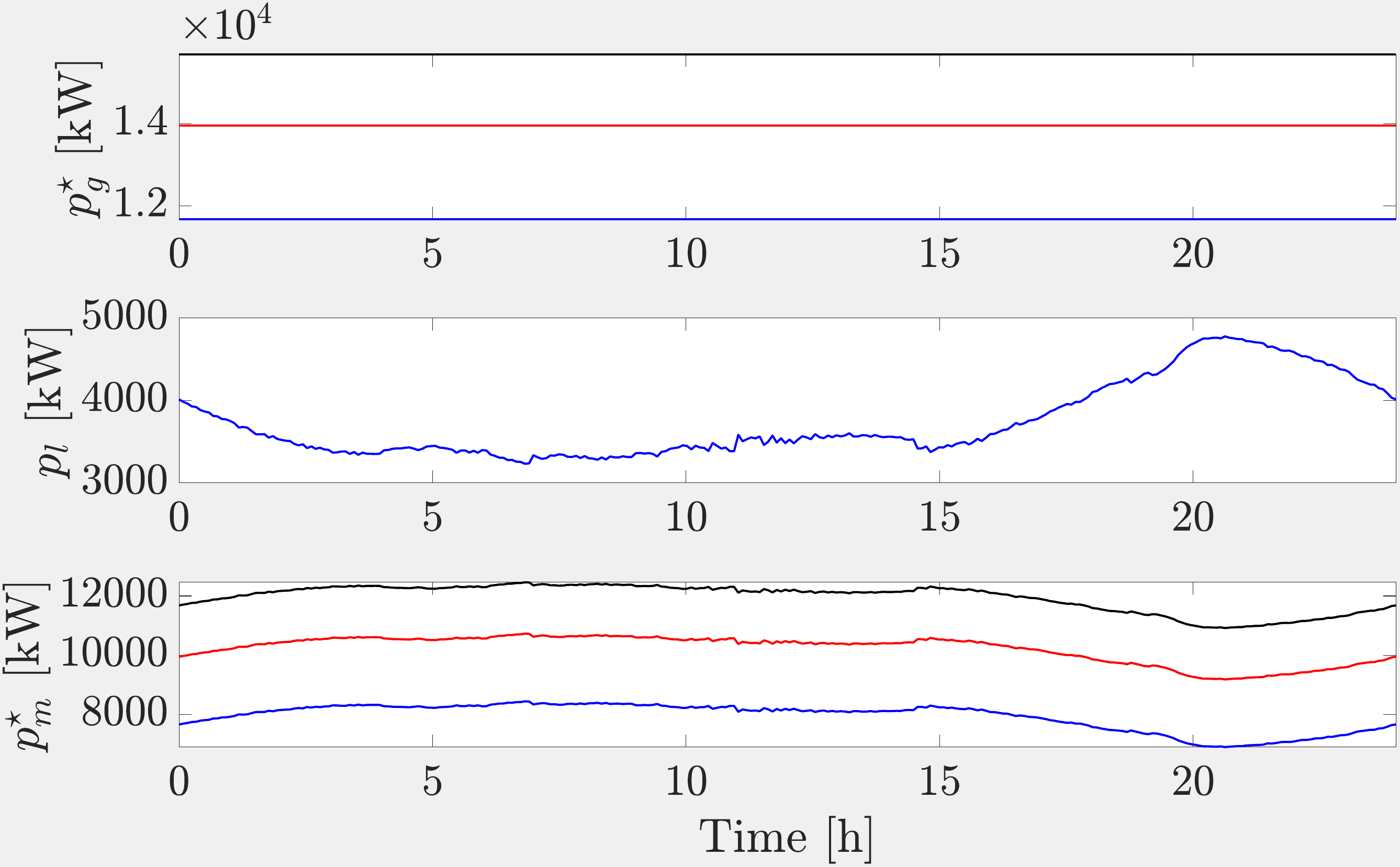}%
            \label{subfig:b}%
        }\\
        \subfloat[July]{%
            \includegraphics[width=.4\linewidth]{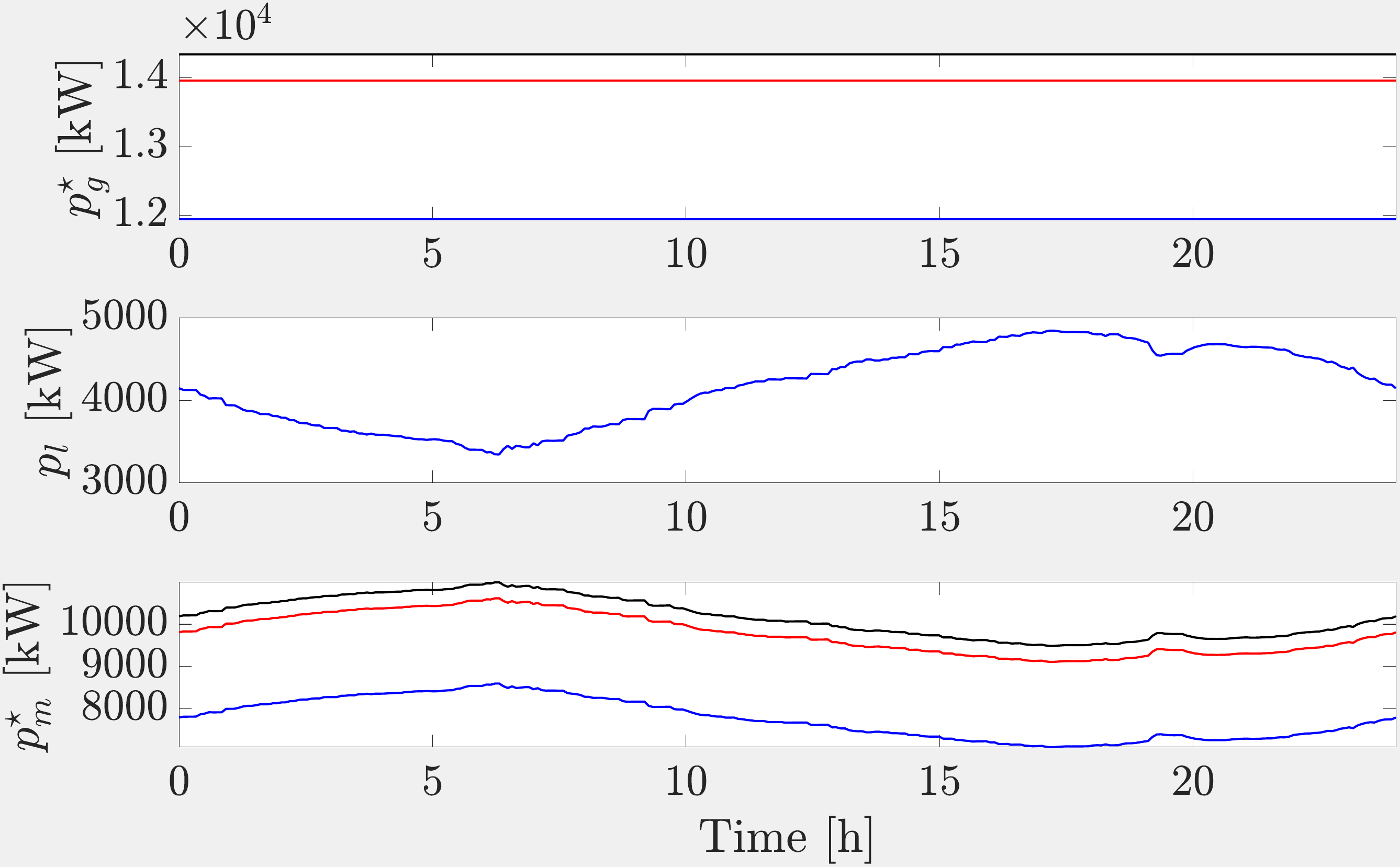}%
            \label{subfig:c}%
        }\hfill
        \subfloat[January]{%
            \includegraphics[width=.4\linewidth]{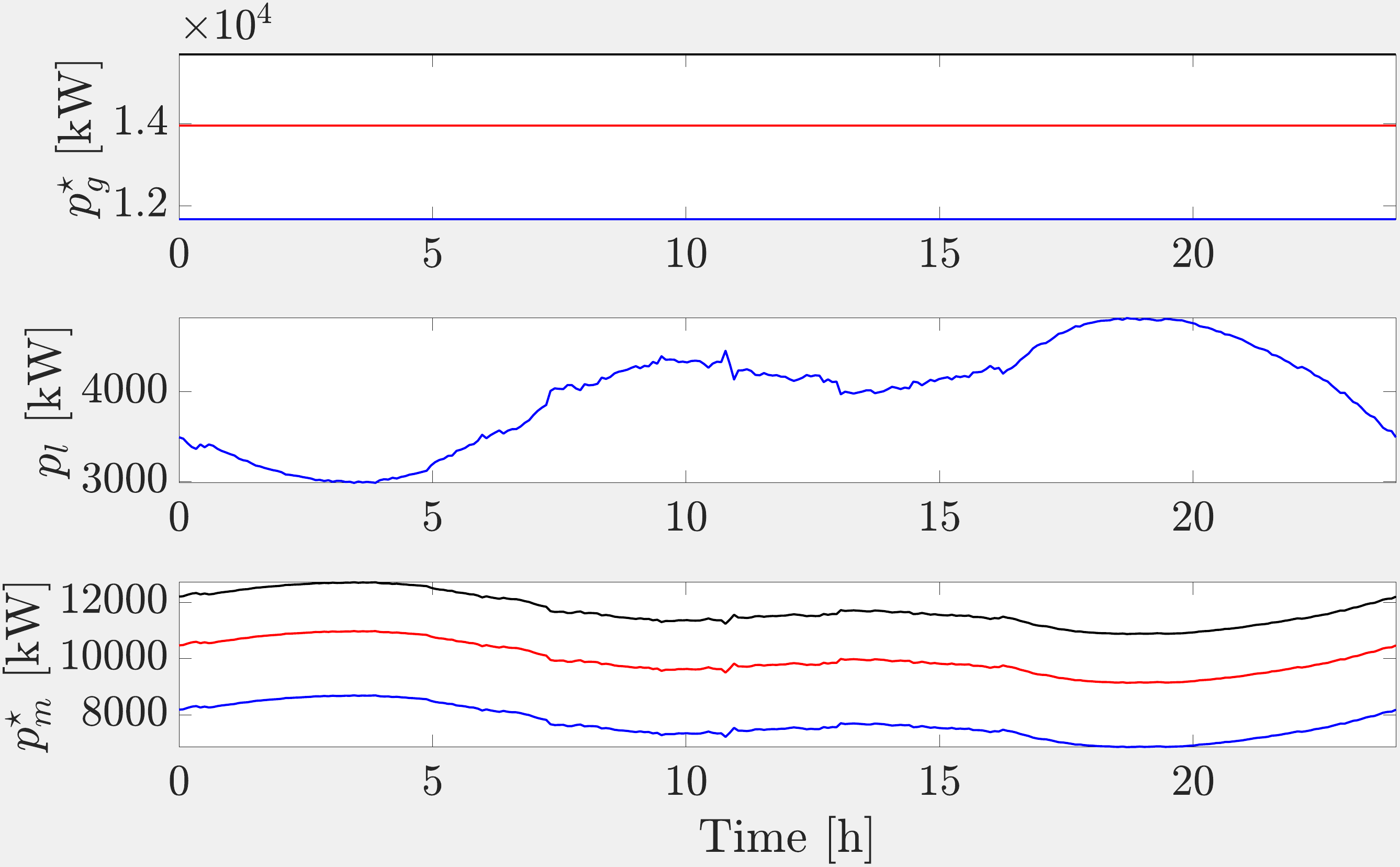}%
            \label{subfig:d}%
        }
        \caption{Each figure represents different load and PV profiles according to different months in a year: (a) October; (b) April; (c) July; (d) January. The top subplot of each figure exhibits the energy generation for that day; the second subplot presents the consumption for that day, and the last subplot presents the operational scheduling of the mining machines. The black line represents machine number ``1'' machine, the blue line shows results for ``2'' machine and the red line represents the ``3'' machine.}
        \label{fig:results}
    \end{figure*}

The results are shown in Fig.~\ref{fig:results}. In the graphs, there are four sampled days representing a typical load behavior over the year. For each figure, the top subplot exhibits the energy generation for that day; the second subplot presents the consumption for that day, and the last subplot presents the operational scheduling of the mining machines. As seen from the results, the generation profiles are constant, clearly leading to an optimal reduction in ramping costs. 

\begin{figure}[htbp]
    \centering
    \includegraphics[width=0.6\linewidth]{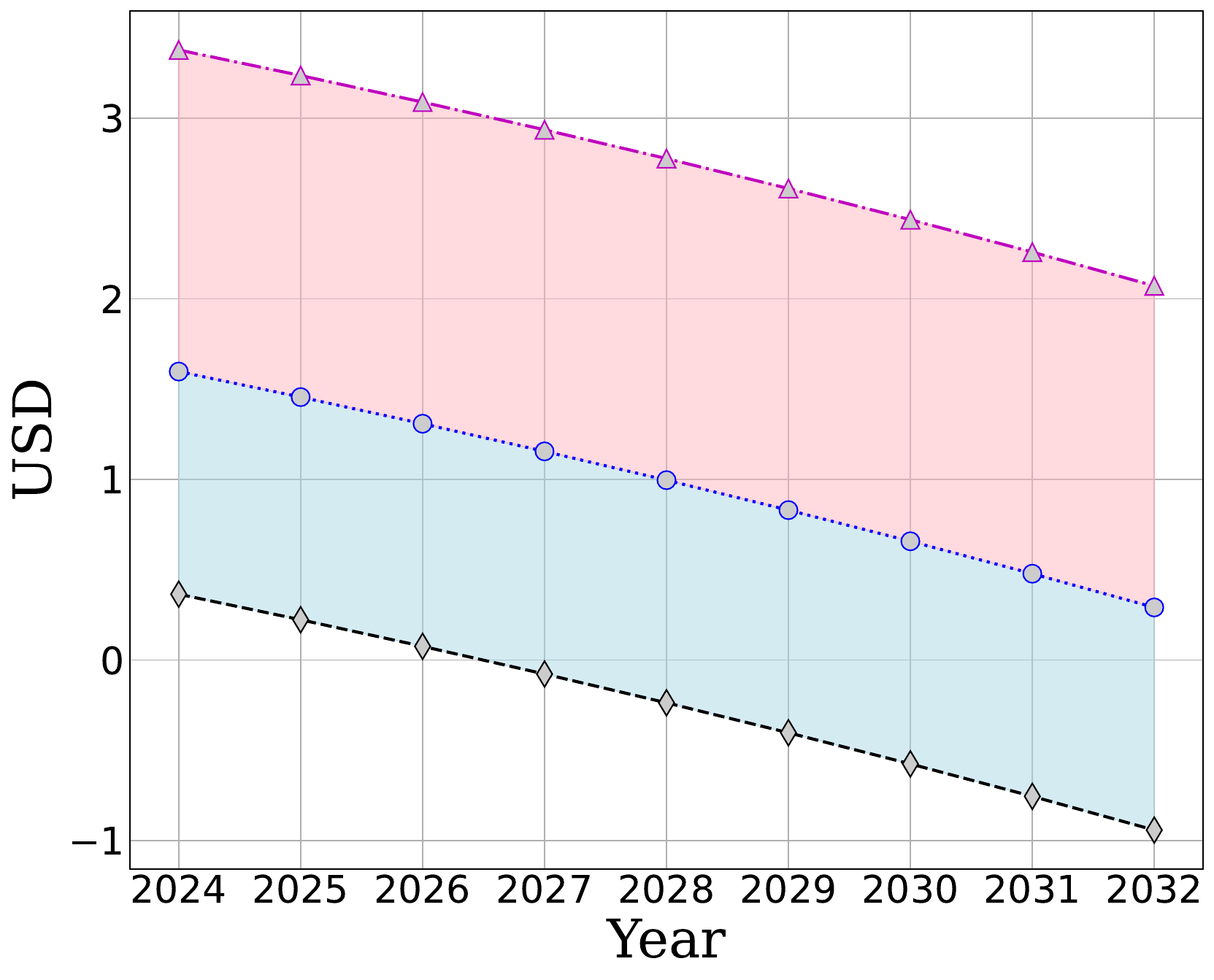}
    \caption{The operator's net profit from Bitcoin mining machines, based on average parameters, over a time horizon of 6 years.}
    \label{fig:net-profit-over-time}
\end{figure}

In \Cref{tab:results}, the generation costs, and revenue from the machine are presented, calculated for 4 months: April, July, and October of 2023, and January of 2024, that represent diverse renewable energy production profiles, and various consumer behaviors patterns during the year. It is clear from the generation patterns that the ramping costs are eliminated if a perfect knowledge of the load profile exists. We assumed 2853 machines of type ``1'', 2175 machines from type ``2'', and 2924 machines from type ``3''.
\begin{table}[htbp]
    \caption{Net profit from mining considering machine costs and electricity costs for single machine over one day per [kW]}
    \label{tab:results}
    
    \begin{tabular}{|p{1cm}|p{1cm}|p{1cm}|p{1.8cm}|p{1.1cm}|} 
    \hline
         Month & Machine & MSRP [\$/day] & Operating costs [\$/day] & Net profit [\$/day] \\ 
         \hline
         April & 1       & 10.14         & 23.77                      &  1.48 \\ 
               & 2       & 7.12          & 31.18                      & 3.76 \\ 
               & 3       & 8.90          & 23.19                      &  2.7 \\ 
         \hline
         July  & 1       & 10.14         & 22.22                      &  1.64 \\ 
               & 2       & 7.12          & 29.14                      & 3.97 \\ 
               & 3       & 8.90          & 21.68                      &  2.92 \\ 
         \hline
         January & 1       & 10.14         & 22.91                      &  1.57 \\ 
                 & 2       & 7.12          & 30.05                      & 3.88 \\ 
                 & 3       & 8.90          & 22.35                      &  2.86 \\ 
         \hline
         October & 1       & 10.14         & 22.06                      &  1.65 \\ 
                 & 2       & 7.12          & 28.93                      & 3.98 \\ 
                 & 3       & 8.90          & 21.52                      &  2.95 \\ 
         \hline
    \end{tabular}
\end{table}
Thus, taking into account the machine price of $7400$ for machine of type ``1'', $5200$ for machine of type ``2'', and $6500$ for machine of type ``3'' (considering machine lifespan of 2 years), the annual revenue from the machines, and the ramping costs reduction is significant. Nonetheless, although this idea might sound attractive, there is a sting in its tail. Looking at the revenues expected over a time horizon of 6 years, it is clear that for machine prices close to machine of type ``1'', the revenue does not account for the purchasing costs and the company will incur financial losses. Figure~\ref{fig:net-profit-over-time} presents this idea. Consequently, until Israel reaches 40\% of renewable energy production, it seems to be beneficial for power plant operators to utilize mining machines for ramping cost reduction, as long as the daily profit from mining are greater than 5\$ and the mining machine price does not exceed 6947\$. As the daily monetary profit from mining, represented by $V$, increases, it allows the system operator to invest more money in the mining machine's purchase, as long as the price of the machine, denoted by $C$, submits to the following:
\begin{equation}
    \frac{C}{2\cdot365} - 8.9 \leq V - 5.
\end{equation}

\bigskip

\section{Conclusion}
In this paper, we examined an extended demand-response program, based on Bitcoin mining machines, for suppressing ramping costs and mitigating related transient effects in the Israeli electrical grid. We analyzed a trend of increasing electricity prices and ramping costs due to the increasing penetration of renewable energy sources, based on historical data taken from ``California ISO''. We formulated and analyzed an extension of unit commitment problem, to improve grid stability and decrees operational costs. Based on this analysis, we suggested a scheduling scheme of Bitcoin mining machines, serving as a flexible load. Following, we investigate the profitability of applying this ancillary service to the ``Noga'' grid operator. We used simulation and the real-world data acquired to verify the proposed demand-response program performance and test its practical limits for reducing the ramping costs and mitigating related transient effects, under changing ratio of energy production from renewable sources. Moreover, in the simulation we examined the profitability of such program for grid operator, taking into account several influential parameters including electricity price, machine price, hashrate and monetary revenue of the machines. We conduct a comparative analysis using the aggregated data to further emphasize the effect of the different properties of these machines, and compare different machine types available in the market and used in cryptocurrency mining farms. Our results suggests that the machine price and ratio of production from renewable sources plays a significant role in determining the profitability of the proposed demand-response program.


\vfill
\end{document}